\documentclass[preprint,showpacs, amsmath, amssymb,superscriptaddress]{revtex4-1}
\usepackage{graphicx}
\usepackage{dcolumn}
\usepackage{bm}
\usepackage{natbib}
\newcommand{\BCPO}{BiCu$_{2}$PO$_6$}

\newcommand{\CuIon}{\ensuremath{\text{Cu}^{\text{2+}}\,}}

\begin{document}
\title{Quasiparticle-continuum level repulsion in a quantum magnet}
\author{K.W. Plumb}
\author{Kyusung Hwang}
\affiliation{Department of Physics and Center for Quantum Materials, University
    of Toronto, Toronto, Ontario M5S 1A7, Canada }
\author{Y. Qiu}
\affiliation{NIST Center for Neutron Research, National Institute of Standards
    and Technology, Gaithersburg, Maryland 20899, USA}
\affiliation{Department of Materials Science and Engineering, University of Maryland, College Park, Maryland 20742, USA}
\author{Leland W. Harriger}
\affiliation{NIST Center for Neutron Research, National Institute of Standards
    and Technology, Gaithersburg, Maryland 20899, USA}
\author{G.~E.~Granroth}
\affiliation{Neutron Data Analysis and Visualization Division, Oak Ridge
National Laboratory, Oak Ridge, Tennessee 37831, USA}
\author{Alexander I. Kolesnikov}
\affiliation{Spallation Neutron Source, Oak Ridge National Laboratory, Oak
    Ridge, Tennessee 37831, USA}
\author{G. J. Shu}
\affiliation{Center for Condensed Matter Sciences, National Taiwan University,
Taipei, 10617 Taiwan}
\author{F.C. Chou}
\affiliation{Center for Condensed Matter Sciences, National Taiwan University,
    Taipei, 10617 Taiwan}
\author{Ch. R\"uegg}
\affiliation{Laboratory for Neutron Scattering and Imaging, Paul Scherrer Institute,
    CH--5232 Villigen, Switzerland}
\affiliation{Department of Quantum Matter Physics, University of Geneva,
    CH--1211 Geneva, Switzerland}
\author{Yong Baek Kim}
\affiliation{Department of Physics and Center for Quantum Materials, University
    of Toronto, Toronto, Ontario M5S 1A7, Canada }
\affiliation{Canadian Institute for Advanced Research/Quantum Materials
    Program, Toronto, Ontario MSG 1Z8, Canada}
\author{Young-June Kim}
\affiliation{Department of Physics and Center for Quantum Materials, University
    of Toronto, Toronto, Ontario M5S 1A7, Canada }


\maketitle
{\bf When the energy eigenvalues of two coupled quantum states approach each 
    other in a certain parameter space, their energy levels repel each other
    and level crossing is avoided. Such level repulsion, or avoided level
    crossing, is commonly used to describe the dispersion relation of 
    quasiparticles in solids. However, little is known about the level 
    repulsion when more than two quasiparticles are present; for example, 
    in an open quantum system where a quasiparticle can spontaneously decay 
    into many particle continuum. Here we show that even in this case level 
    repulsion exists between a long-lived quasiparticle state and a continuum.  
    In our fine resolution neutron spectroscopy study of magnetic 
    quasiparticles in a frustrated quantum magnet \BCPO{}, we observe a 
    renormalization of quasiparticle dispersion relation due to the presence of 
    the continuum of multi-quasiparticle states. Our results have a broad
    implication for understanding open quantum systems described by 
    non-hermitian Hamiltonian.}

A fundamental concept in condensed matter physics is the idea that strongly
interacting atomic systems can be treated as a collection of weakly interacting
and long-lived quasiparticles. Within a quasiparticle picture, complex
collective excited states in a many body system are described in terms of
effective elementary excitations. The quanta of these excitations carry a
definite momentum and energy and are termed quasiparticles. Magnetic insulators
containing localized $S\!=\!1/2$ magnetic moments and having valence bond solid
ground states are ideal systems in which to study bosonic quasiparticles in an
interacting quantum many body system~\cite{Giamarchi:08}. The elementary
magnetic excitations in these materials  are triply-degenerate $S\!=\!1$
quasiparticles called triplons, and their momentum and energy resolved dynamics
can be probed directly though inelastic neutron scattering.

In particular, when the system's Hamiltonian has an interaction term coupling
single and multi-particle states, the single quasiparticles may decay
into the continuum of multiparticle states \cite{Stone:06, Zhitomirsky:2013}.
Therefore, an ensemble of quasiparticles as may be realized in a quantum magnet
is a good example of an open quantum system, in which particle number is not
conserved. \footnote{The term, {\em open quantum system}, is often used in a different context, referring to a system that interacts with an environment, and follows stochastic dynamics. [See for example, H.-P. Breuer and F. Petruccione, {\it The theory of open quantum systems} (Oxford Univ. Press, Clarendon, 2007).] In this article, however, we are only concerned with a system described by non-hermitian Hamiltonian, in which particles can decay.}
In such a system, the energy eigenvalues are in general complex,
and the particles have a finite lifetime, which is given by the imaginary part
of the complex eigenvalue \cite{Rotter:2009}. It was shown that even in the case of an
open quantum system, the avoided level crossing could occur in the complex
plane \cite{Okolowicz:2003}.

Despite broad interest in interacting open quantum systems, experimentally
realizing an ideal condition to study the interaction between a quasiparticle
and a multi-particle continuum turns out to be extremely difficult. Here we
show that the quantum magnet \BCPO{} presents a rare model system that allows
one to study quasiparticle level repulsion in the complex plane. This phenomena
arises due to the presence of the very large anisotropic exchange
interactions in \BCPO{}, originating from spin-orbit coupling (SOC).  These
anisotropic interactions play a dual role in \BCPO{}. First, they break the
degeneracy of the triplet excitations, making the triplon dispersion relation
non-degenerate in phase space. In addition, the anisotropic exchange
interaction is responsible for the strong anharmonicity which couples the
single and multi-triplon excitations. As a result, the triplon lifetime may be
reduced in the region of phase space where (single) triplon dispersion overlaps
with multi-triplon continuum. This decay process may in fact be so strong that
the quasiparticle description ceases to be valid within the continuum.
Furthermore, we show that the dispersion relation of the triplons is strongly
renormalized near the boundary of the multi-triplon continuum due to the level
repulsion between the single quasiparticle and the continuum.

In the following, we first present inelastic neutron scattering (INS)
measurements of the full triplon dispersion in \BCPO{}, which reveal a rich
excitation spectrum including a multi-triplon continuum of scattering.
Analysis of the triplon excitation spectrum using bond-operator theory enables
us to determine the magnetic Hamiltonian of \BCPO{} accurately. We find that
strong spin-orbit coupling plays an essential role in this compound
through substantial symmetric and antisymmetric anisotropic interactions. We
will finally discuss the interaction of the triplon quasiparticles with the
continuum which manifests as a drastic renormalization of the quasiparticle
spectra and ultimately spontaneous quasiparticle decay.

\section*{Valence Bond Solid}
\begin{figure*}
    \includegraphics[]{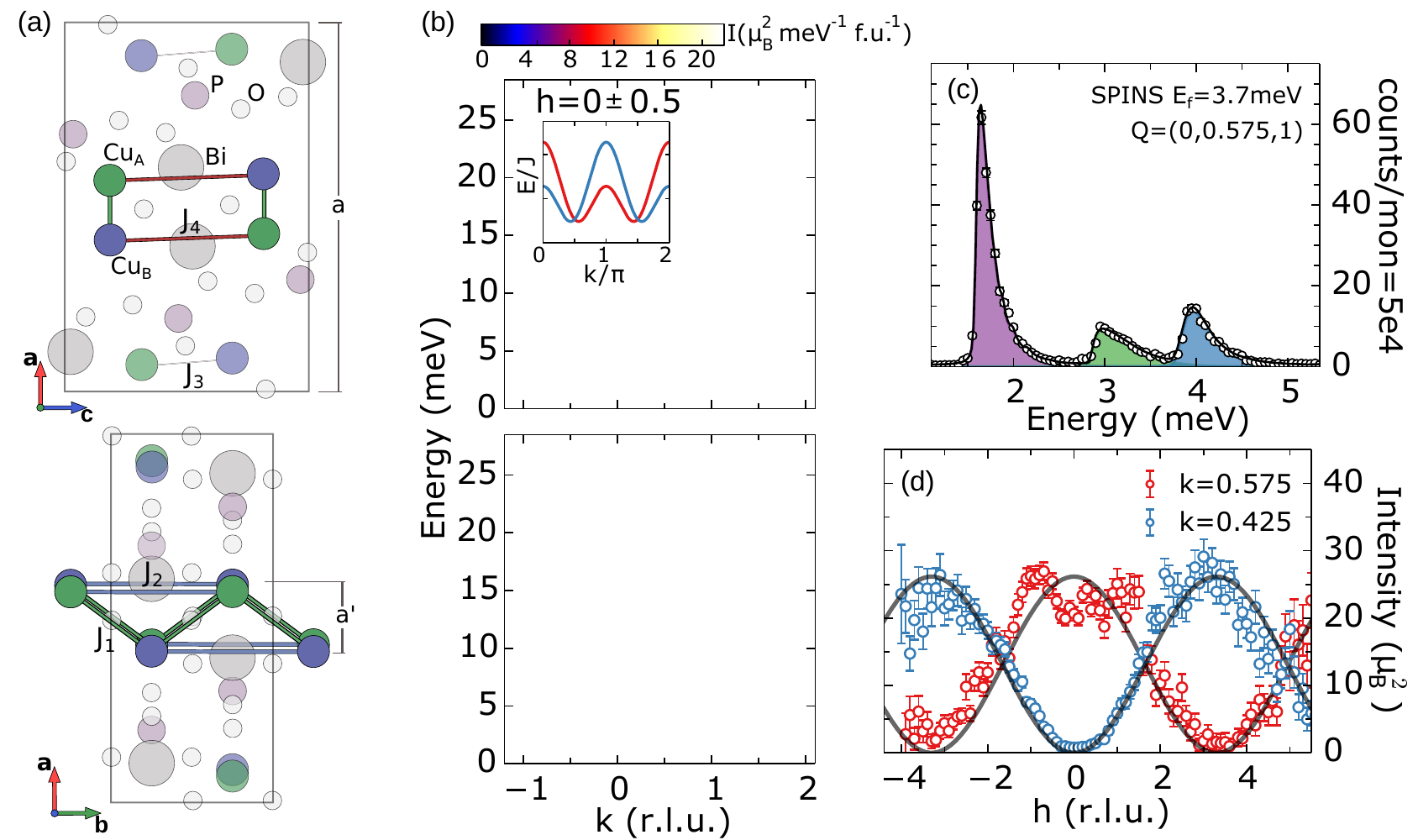}
        \caption{\label{fig:SQW} (a) The orthorhombic crystal structure of
            \BCPO{} contains zig-zag chains of \CuIon ions running parallel to
            the b-axis. The coupled chains can be viewed as bi-layers of \CuIon
            with spacing $\rm a^{\prime}\!  =\!0.162 a$.  Along the chains
            neighboring \CuIon interact via competing nearest-neighbor (NN) and
            next-nearest-neighbor (NNN) antiferromagnetic exchange terms $J_1$
            and $J_2$.  The chains are coupled along the c-axis by $J_4$ to
            form a frustrated two-leg ladder, with an additional weak
            interladder exchange $J_3$. (b) Energy-momentum slice of the
            inelastic neutron scattering intensity at T=4~K. Data have been
            corrected for the isotropic \CuIon form-factor \cite{Brown:06} and
            intensities placed into absolute units using the incoherent
            scattering from a vanadium standard.  Inset is a schematic
            illustration of the dispersion of each mode. (c) Constant momentum
            transfer scan at the incommensurate wavevector Q = (0,0.575,1)
            collected on the SPINS spectrometer at T=75~mK. Solid line is a fit
            to a resolution convolved model cross-section and filled areas are
            contributions from each mode.  (d) Constant energy transfer cuts
            along two-dimensional rods of scattering at the $k=0.5\pm0.075$
            positions integrated over $E\!=\!2\pm0.5$~meV.  Solid lines are the
            bilayer structure factor as described in the text.  Error bars
            represent one standard deviation.}
\end{figure*}

The orthorhombic crystal structure of \BCPO{} is shown in Fig.~\ref{fig:SQW}
(a); the structure contains zig-zag chains of \CuIon ions running parallel to the
b-axis. Magnetic interactions along the chains are frustrated because of a
competition between the nearest-neighbor (NN) and next-nearest-neighbor (NNN)
antiferromagnetic exchange terms $J_1$ and $J_2$. The frustrated chains are
coupled strongly along the c-axis by antiferromagnetic coupling $J_4$ to form a
two-leg ladder. As a result of the strong antiferromagnetic coupling $J_4$, two
spins on each rung can form singlets, and the ground state is described as an
array of singlets termed a valence bond solid. The elementary excitations in
this case are triplets that can propagate along the chain direction due to
$J_1$ and $J_2$. Within a single (ladder) bilayer, there are two
crystallographically inequivalent copper sites (Cu$_A$ and Cu$_B$ as shown in
Fig.~\ref{fig:SQW} (a)). This results in the breaking of inversion symmetry
across all magnetic bonds, and consequently anisotropic interactions are permitted in
the magnetic Hamiltonian, as will be discussed later.

The dispersion relation, which contains essential information regarding the
triplon dynamics such as the effective mass and velocity, is revealed directly
by inelastic neutron scattering. Before discussing our neutron scattering
measurements, it is helpful to briefly review the expected excitation
spectrum of the frustrated two-leg ladder as realized by \BCPO{}, first
ignoring any anisotropic interactions. In the strong coupling limit of
$J_1,J_2\!\ll\!J_4$, the expected excitation spectrum is schematically
illustrated in the
inset of Fig.~\ref{fig:SQW} (b)\cite{Tsirlin:10,Lavarelo:11}. The
dispersion has a distinct W shape with minimum at an incommensurate wavevector.
The incommensurate minimum of the gapped spectrum is the manifestation of the
magnetic frustration in \BCPO{}; since \BCPO{} contains
two singlet dimers per unit cell, there are two separate bands of
triplons.  The bands become degenerate at the point $k= 0.5$ and are related by
a simple folding of the zone with the minima of each band appearing at
incommensurate wavevectors $k = 0.5\pm\delta$.  Anisotropic interactions
entering the magnetic Hamiltonian may then further split the degeneracy of each
triplon band.

An overview of the zero field INS measurements is presented in
Fig.~\ref{fig:SQW} (b).  The gapped, W-shaped, dispersion of each branch is
clearly visible at both $h\!=\!0$ and $3$ with a bandwidth of 25~meV, and
incommensurate minima at $k=0.575$ and $k=0.425$ in (a) and (b) respectively.
We have not observed any dispersion along the $\mathbf{h}$-direction confirming
the weak inter-bilayer coupling; however, intensities are strongly modulated
with momentum transfers along $h$. This effect is most clearly shown in
Fig.~\ref{fig:SQW} (d) where the intensity along the two dimensional rods of
scattering, at $k=0.5\pm0.075$ positions, is plotted. The modulation arises from the bilayer structure factor: interference between the scattering from the two layers within a bilayer. For weak inter-bilayer
coupling the bilayer structure factor can be written simply in the following
form $ B^{i}(\mathbf{Q})\! =\!
A\cos{\left(\frac{1}{2}\mathbf{Q}\!\cdot\!a^{\prime}\mathbf{\hat{h}}+\phi^{i}\right)}$,
where $a^{\prime}$ is the intrabilayer spacing, shown in Fig.~\ref{fig:SQW}
(a), $\mathbf{\hat{h}}$ is a unit vector directed along $\mathbf{h}$,
$\phi^{i}$ is a phase factor for the mode indexed by $i$, and $A$ is an
arbitrary amplitude factor.  Solid lines in
Fig.~\ref{fig:SQW} (d) are the bilayer structure factor with the known
intrabilayer spacing for \BCPO{} of $a^{\prime}\!=\!0.162a$, and 
$\phi^{i}\!=\!0$
and $\pi/2$ for the modes with minima at $k=0.575$ and $k=0.425$ respectively.

What makes \BCPO{} unique among valence bond solids is the presence of strong
anisotropic interactions that qualitatively alter the nature of the triplons.
The evidence for anisotropic interactions in \BCPO{} is first borne out by
high resolution measurements around the incommensurate wavevector at
$\mathbf{Q}\!=\!(0,0.575,1)$ shown in Fig.~\ref{fig:SQW} (c). These reveal that
anisotropies in \BCPO{} completely split the degeneracy of each primary branch
such that three distinct modes are observed.  Note that the intensity
modulation of the primary modes by the bi-layer structure factor enables the
independent probing of each mode. Anisotropy splitting is significant, with the
minima of each mode corresponding to gap values of $\Delta_1\!=\!1.67(2)$~meV,
$\Delta_2\!=\!2.85(5)$~meV, and $\Delta_3\!=\!3.90(5)$~meV.

\begin{figure*}
        \includegraphics{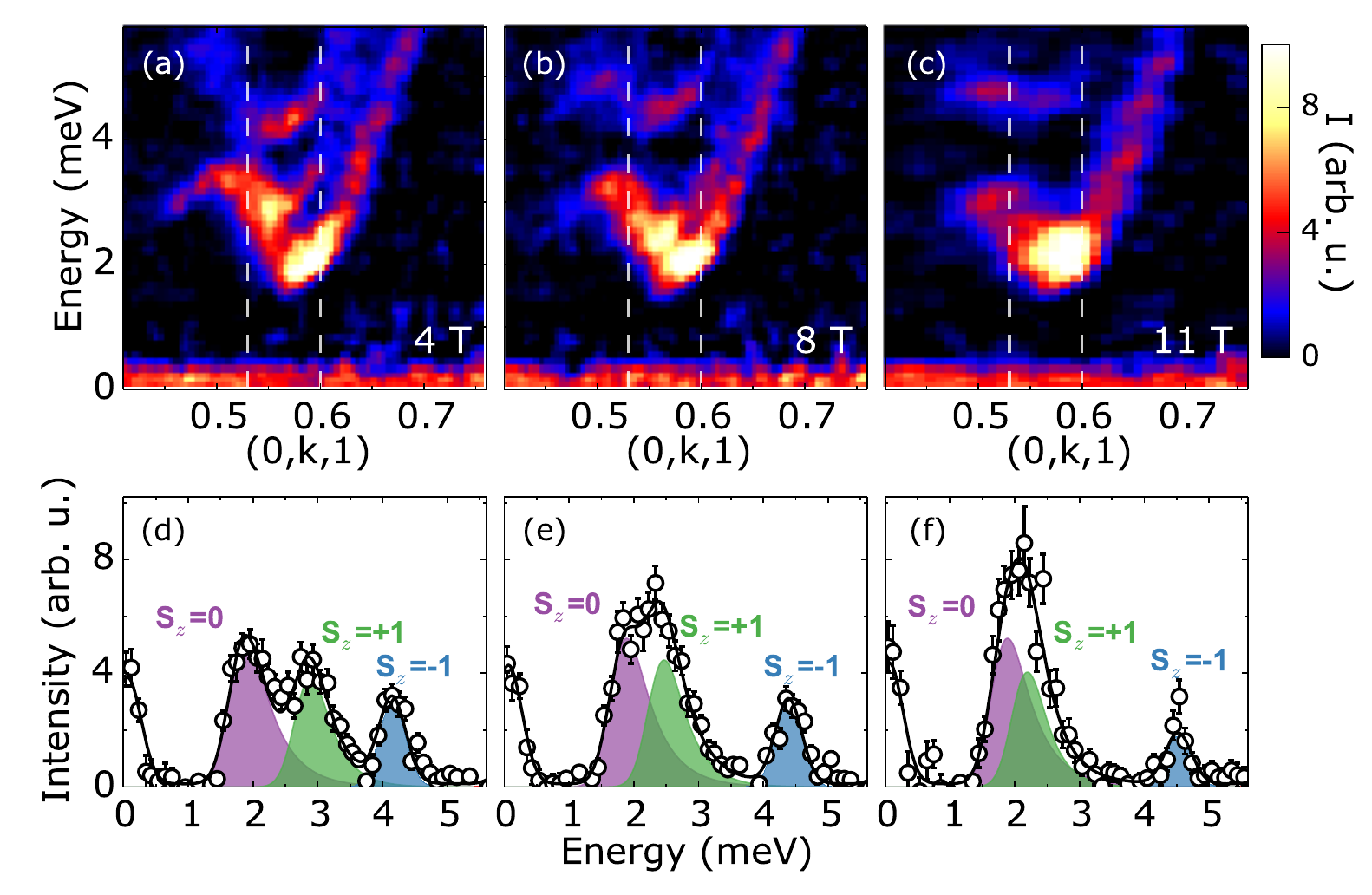}
        \caption{\label{fig:Hdep} Magnetic field dependence of dynamic magnetic
            correlations in \BCPO{}. (a)-(c) False color maps of the INS
            intensity measured on DCS at T=100~mK with magnetic field applied
            along the crystallographic $\mathbf{a}$-axis.  (d)-(f) Constant
            momentum transfer cuts integrated over $\rm
            k\!=\!0.565\pm0.035$~r.l.u.\@ and $\rm l\!=\!1\pm0.05$~r.l.u.\@,
            integration ranges are represented by dashed white lines in
            (a)-(c).  Solid lines are fit to asymmetric Gaussian functions and
            filled areas show the contribution from each mode.  Asymmetric line
            shapes result from the steep dispersion and extended momentum
            integration range.  Error bars represent one standard deviation.}
\end{figure*}

The quantum states of each mode can be further explored via INS
measurements performed with applied magnetic fields. The neutron intensities
for applied fields of 4, 8, and 11~T are plotted in Fig.~\ref{fig:Hdep}. No
further splitting of the modes was observed indicating that anisotropic
interactions in the Hamiltonian have completely lifted the SU(2) spin rotation
symmetry~\footnote{In this work, measurements in applied fields are all for the
    very narrow region $h\!=\!0\pm0.2$ r.l.u.\ because the magnet aperture
    limited the detection of neutrons with momentum transfer out of the
    horizontal scattering plane}. Constant momentum-transfer cuts around the 
incommensurate wavevector, Fig.~\ref{fig:Hdep} (d)-(f), reveal an anomalous 
Zeeman behaviour. Rather than splitting into the conventional ordering in 
energy of $S_z\! =\! \{+1,0,-1\}$ \cite{Matsumoto:2002,Giamarchi:08}, the 
lowest energy mode exhibits negligible field dependence and is assigned a $S_z 
= 0$ quantum state, while the two higher energy modes have the Zeeman character 
of $S_z\!=\!+1$ and $S_z\!=\!-1$ respectively. (See the field dependence 
plotted in Fig.~3c.) \footnote{ Of course, in the presence of anisotropic 
    interactions the singlet and triplet wave functions are mixed and $S_z$ is 
    no longer a good quantum number. Here we assign each mode a pseudo $S_z$ 
    based on on its Zeeman energy as this provides a convenient labelling 
    scheme.}

\section*{Non-interacting triplons}
The complete dispersion extracted from INS measurements, combining the data for
$h=0$ and $h=3$, is plotted in Fig.~\ref{fig:Dispersion}. There are six triplon
modes, two bands arise from two inequivalent dimers per unit cell which are
each further split into three non-degenerate modes by anisotropic interactions.

To understand the spin dynamics in \BCPO{} we consider a generic spin
Hamiltonian with Heisenberg $J_{ij}$, as well as antisymmetric
$\mathbf{D}_{ij}$ and symmetric anisotropic $\Gamma_{ij}^{\mu\nu}$ interactions
\begin{align}
    \label{eq:Hamiltonian}
    \mathcal{H}\!=&\!\sum_{i>j}\!\left(J_{ij}\mathbf{S}_i\!\cdot\!\mathbf{S}_j\!
            +\!\mathbf{D}_{ij}\!\cdot\!\mathbf{S}_i\!\times\!\mathbf{S}_j\!+\!\Gamma_{ij}^{\mu\nu}S^{\mu}_iS^{\nu}_j\right)\\\nonumber
        &-g\mu_B\mathbf{H}\cdot\!\sum_i\!\mathbf{S}_i,
\end{align}
in which the symmetric anisotropy term is constrained by the relation
\begin{equation}
    \label{eq:Gamma}
    \Gamma_{ij}^{\mu\nu}\!=\!\frac{D_{ij}^{\mu}D_{ij}^{\nu}}{2J_{ij}} -
    \frac{\delta^{\mu\nu}\mathbf{D}_{i,j}^2}{4J_{i,j}}.
\end{equation}
Both the antisymmetric Dzyaloshinskii-Moriya (DM) and symmetric ($\Gamma$)
anisotropic exchange terms originate from spin-orbit
coupling \cite{Dzyaloshinsky:58,Moriya:60, Shekhtman:92, Yildirim:95}. While
the symmetric anisotropy term is the smaller of the two, and is often
neglected, it can have pronounced effects on the magnetic ground state and
excitation spectrum \cite{Zheludev:98}. Employing a quadratic (non-interacting)
bond operator theory (BOT) \cite{Sachdev:1990,Gopalan:1994,Matsumoto:2004}, for
the valence-bond ordered ground state with valence bonds on $J_4$ links, we
have found that the INS data is best described with the following coupling
constants: $J_1\!=\!J_2\!=\!J_4\!=\!8$~meV, $J_3\!=\!1.6$~meV,
$D_1^a\!=\!0.6J_1$, $D_1^b\!=\!0.45J_1$, $\Gamma_1^{aa}\!=\!0.039J_1$,
$\Gamma_1^{bb}\!=\!-0.039J_1$, and
$\Gamma_1^{ab}\!=\!\Gamma_1^{ba}\!=\!0.135J_1$~\footnote{See supplementary information
for details of the bond operator theory.}; the calculated triplon
dispersion is plotted in Fig.~\ref{fig:Dispersion} (a) and (b).  At quadratic
order, the BOT captures important details of the low energy spectra including
the slight shift of incommensurate minima between each branch,
as well as the overall bandwidth of the excitations. Importantly, this calculation appropriately describes the anomalous Zeeman splitting
plotted in Fig.~\ref{fig:Dispersion} (c).  Furthermore, extending the BOT \footnote{See supplementary information
    for the calculated Zeeman behaviors of
    excitations for $H \parallel \mathbf{a},\mathbf{b}, \mathbf{c}$.} to
determine the field dependence of each mode for fields applied along the
$\mathbf{b}$ and $\mathbf{c}$ directions correctly predicts the hierarchy of
critical fields measured previously
$H_c^a\!>\!H_c^b\!>H_c^c$.\cite{Kohama:12} The
quadratic BOT is essentially a mean-field expansion and so
is likely to overestimate the coupling constants.  However, even within the
mean field estimation, it is remarkable that such large anisotropic
interactions are required to describe magnetic excitations in \BCPO{}.

While the BOT describes the overall features of the measured triplon
dispersion, this quadratic theory fails to capture some very distinct features
of the spectrum, including a bending of the triplon modes around
k$_c\approx1\!\pm\!0.26$ in Fig.~\ref{fig:Dispersion} (a).  Here the single
triplon dispersions are strongly renormalized, bending away from the quadratic
dispersion, in an avoided crossing with a multi-triplon continuum. In addition,
the dispersion abruptly stops beyond the critical wavevector of $k_c \approx
0.75$. As we will discuss below, these dynamics ultimately arise as a
consequence of the anharmonic magnetic interactions which couple single triplon
quasiparticles with multi-triplon continuum states.

\section*{Multiparticle continuum and level repulsion}
The presence of large  anisotropic exchange interactions has  dramatic
implications on the behaviour of triplons in this system. In contrast to
isotropic quantum magnets where the triplon dynamics are typically well
described in a harmonic expansion, the anisotropic exchange interactions in
\BCPO{} produce significant anharmonic (cubic order in bond operators) couplings appearing as non-particle
conserving terms in the bond-operator Hamiltonian. This anharmonicity modifies
the triplon dispersion relation in \BCPO{} qualitatively.  In
figure~\ref{fig:TermPoints} (b) the quantity $\hbar\omega S(\mathbf{Q},\omega)$
is represented in the false color map.  $\hbar\omega S(\mathbf{Q},\omega)$
highlights the effects of multi-triplon interactions, including an avoided
crossing and extinction of the lowest bands around $k=0.8$ and a continuum of
scattering at high energies around $k=1$.

\begin{figure*}
    \centering
        \includegraphics{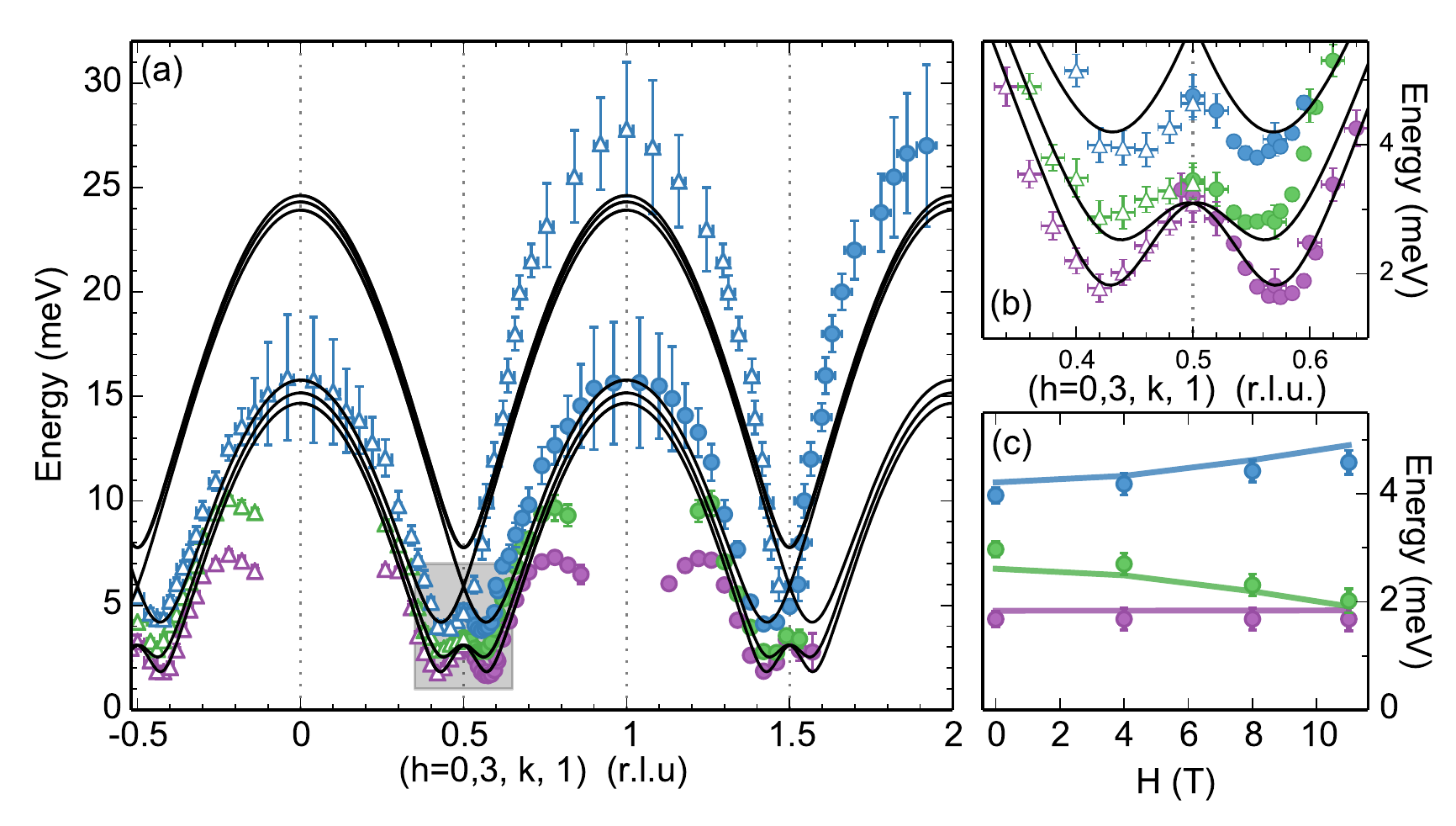}
        \caption{\label{fig:Dispersion} (a) Dispersion of triplon excitations
            in \BCPO\!\!. Points extracted from fits to constant energy and
            momentum transfer scans from SEQUOIA and SPINS data sets. Filled
            circles correspond to $h = 0$, and open triangles correspond to $h
            = 3$.  Solid lines are the dispersion calculated with a quadratic
            bond-operator theory as described in the text.  (b) Detailed view
            around the point $k\!=\!0.5\pm0.075$, indicated by the black shaded
            region in (a).  (c) Measured energy of each mode at $\rm
            Q\!=\!(0,0.575,1)$ as a function of applied magnetic field.  Lines
            in (c) are the corresponding calculated Zeeman energies from the
            bond-operator theory.}
\end{figure*}

\begin{figure*}[htbp!]
    \centering
        \includegraphics{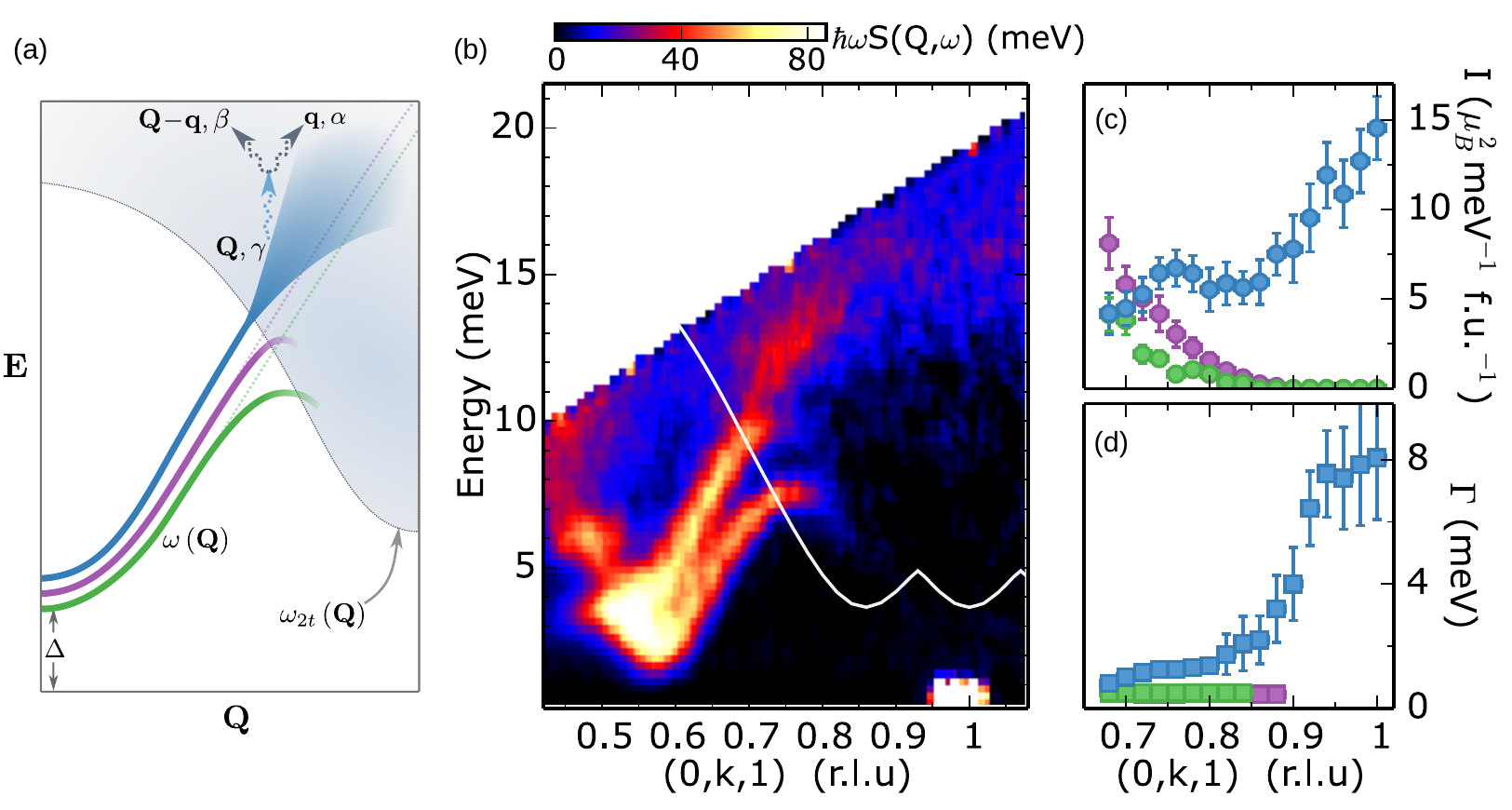}
        \caption[]{\label{fig:TermPoints} Level repulsion and termination of
            quasiparticle spectrum in \BCPO{}.(a) Schematic illustration of the
            renormalization and termination of different single-triplon
            branches as they approach the multi-triplon continuum. Within the
            continuum a single triplon with moementum $\mathbf{Q}$ and spin
            quantum number
            $\gamma$ can decay into two triplons with momentum $\mathbf{q}$ and
            $\mathbf{Q}\!-\!\mathbf{q}$.  (b) $\hbar \omega
            S(\mathbf{Q},\omega)$ highlighting the avoided crossing of the
            triplon branches with the continuum scattering around Q$_c$.  Solid
            white line is the lower bound for two-triplon scattering calculated
            from the non-interacting bond-operator theory. Momentum dependence
            of the spectral weight contained in each branch (c) and full width
            half maximum (d).}
\end{figure*}

In addition to exciting a single-triplon quasiparticle, a neutron can create
two or more triplon excitations simultaneously. For example, a neutron with
momentum $\mathbf{Q}$  can create two triplons with momentum $\mathbf{q}$ and
$\mathbf{Q}\!-\!\mathbf{q}$. These two-triplon excitations form a continuum
with a lower bound determined by conservation of momentum and energy
$\omega_{2t}\left(\mathbf{Q}\right)\!  = \rm min_{\mathbf{q}}\lbrace\,
\omega\left(\mathbf{q}\right) \!  + \!  \omega\left(\mathbf{Q}\!  -
    \!\mathbf{q}\right)\rbrace$, where $\omega\left(\mathbf{q}\right)$ is the
single-triplon dispersion.  This lower boundary for two-triplon
scattering, as determined from the quadratic dispersion, approximately
delineates the region of continuum scattering shown in
Fig.~\ref{fig:TermPoints} (b). It is within this kinematic bound that the
qualitative effects of anharmonic interactions in \BCPO{} are most apparent.
Here the non-particle conserving terms offer decay channels for single triplon
excitations.  A physical consequence of this is that the triplon lifetime is
significantly reduced, even in the absence of any thermal fluctuations. The
effect manifests in a neutron scattering experiment as a strong  damping of the
quasiparticle peak. In Figs.~\ref{fig:TermPoints} (c) and (d) the momentum
dependent intensity and linewidth of each mode around $k=0.8$ are shown,
highlighting the different behaviour of each mode in this region.  While we
observe a continuous increase in the  damping of the highest energy mode as it
smoothly merges into the continuum, a much more dramatic effect is apparent in
the two low energy triplon branches. The single
triplon dispersions for these branches are strongly renormalized by interactions with the
continuum, bending away from the quadratic dispersion, in an avoided crossing
as shown in Fig.~\ref{fig:Dispersion} (a). In addition, these branches remain
resolution limited in energy over all wave vectors, but terminate abruptly upon
entering the continuum.  This is a spectacular example of a spontaneous
quasiparticle breakdown, where the decay channels are so effective that an
appropriate description of the system in terms of quasiparticles does not exist
\cite{Zhitomirsky:2006,Zhitomirsky:2013}.

While detailed observations of triplon dynamics have been made in the past
\cite{Xu:00, Stone:01, Notbohm:07}, clear examples of the
spontaneous breakdown of a triplon spectrum as observed here are rare.  In the 
two-dimensional correlated singlet material piperazinium hexachlorodicuprate 
(PHCC), a well defined triplon peak in the excitation spectrum was observed to 
abruptly merge with a continuum and vanish beyond a threshold momentum 
\cite{Stone:06}. Somewhat different phenomena were observed in the 
organometallic two-leg ladder
compound IPA-CuCl$_3$, in which the single triplon dispersion abruptly terminated
beyond a critical wavevector, without damping nor two triplon continuum
\cite{Masuda:06}. The decay process we observe in \BCPO{} is unique, as each
triplon branch exhibits different decay behavior, as illustrated schematically
in Fig.~4a. The highest energy mode does not bend, but merges smoothly with and 
decays into the continuum, in contrast to the behavior of the two lower energy 
branches. Since each single particle mode is associated with a different 
effective spin quantum number, this behaviour may be associated with branch 
dependent selection rules in the decay processes. Such an explicit 
manifestation of spin dependent quasiparticle interactions has not been 
observed before.

We would like to emphasize that the underlying magnetic interactions in \BCPO{}
satisfies two crucial conditions for realizing stong coupling between the
single and multiparticle states. First, the large bandwidth of triplon
excitations relative to the gap energy at the incommensurate minima results in
a large overlap between the single-triplon and continuum states. This overlap
ensures that the kinematic conditions for the decay of a single-triplon are
satisfied over a large region of phase space. Second, strong anharmonic
interactions provide channels for coupling single and multi triplon states
\cite{Zhitomirsky:2006,Zhitomirsky:2013}.  The existence of magnetic
interactions which couple single and multi-particle states are not always
guaranteed.  In \BCPO{}, the Heisenberg exchange terms, $J_1$, $J_2$, and $J_4$
cancel at cubic order in an interacting bond operator theory and do not
contribute to the spontaneous decay.  It is the DM interactions which appear as
the strongest anharmonic terms and, thus, are responsible for the spontaneous
decay of single particle states into multiparticle ones.\footnote{A full 
    theoretical treatment of triplon interactions is beyond the scope of this 
    article, but such interactions will renormalize the DM parameters to 
    somewhat smaller values. Nevertheless substantial $D$ and $\Gamma$ terms 
    are essential for fitting the magnetic excitation spectra.}

In summary, we have mapped the quasiparticle excitation spectra in the quantum
magnet \BCPO{} through comprehensive INS measurements.  The low energy triplon
excitations are captured by a quadratic bond operator theory for the valence
bond solid and we find that very large anisotropic interactions are necessary 
to describe the excitation spectrum.  These
anisotropic couplings appear as anharmonic, non particle conserving, terms in
the bond operator Hamiltonian and manifest in a complete termination of the
quasiparticle spectra beyond a critical wavevector. Strong hybridization
between the lowest triplon branches and higher order continuum scattering in
the neighbourhood of the critical wavevector results in a renormalization and
avoided crossing in this open quantum system. Perhaps the most important
feature of the excitation spectrum in \BCPO{} is this selective hybridization,
renormalization, and termination of the two lowest branches, distinct
from the smooth merging of the highest energy branch into a continuum.
Further theoretical investigation of interacting triplons could shed light on 
the origin of the observed unusual decay behaviour.

{\bf Acknowledgments:} We would also like to thank G\"otz Uhrig, Oleg Tchernyshyov
and Se Kwon Kim for helpful discussions.  This research was supported by NSERC
of Canada, Canada Foundation for innovation, Canada Research Chairs program,
and Centre for Quantum materials at the University of Toronto.  Work at ORNL
was sponsored by the Division of Scientific User Facilities, Office of Basic
Energy Science, US department of Energy (DOE).  Work at NIST utilized
facilities supported in part by the National Science Foundation under Agreement
No. DMR-0944772.

{\bf Methods:} All measurements used the same 4.5~g single crystal as previous
studies \cite{Plumb:2013}. Magnetic excitations in \BCPO{} were mapped through
INS measurements performed on a number of instruments. High energy time of
flight neutron scattering measurements where carried out on the SEQUOIA
spectrometer at the Spallation Neutron source covering the full dynamic range
of excitations in \BCPO{} with high energy resolution $\Delta E\!\sim\!0.8$~meV
at the elastic line.  Measurements on SEQUOIA were performed with a fixed
incident neutron energy of E$_i$= 40~meV and the fine resolution Fermi-chopper
(FC$_2$)\cite{Granroth:2010} rotating at 360~Hz.  The sample was mounted with
the (h,k,0) plane lying in the horizontal scattering plane of the instrument
and (h,0,0) initially aligned along the incident neutron wavevector
$\mathbf{k}_i$.  In order to map the complete dynamic structure factor
$S(\mathbf{Q},\omega)$ the sample was rotated through 180$^{\circ}$ in
0.5$^{\circ}$ steps. All measurements on SEQUOIA were performed with the sample
held at a temperature of 4~K. Another set of high resolution measurements where
conducted on the SPINS cold triple axis spectrometer at the NIST center for
neutron research. Here the sample was mounted in the (0,k,l) scattering plane
and all measurements used a fixed final energy of E$_f$ = 3.7~meV employing a
vertically focusing PG monochrometer, a flat PG analyser, and a BeO filter
between the sample and analyser. The spectrometer collimation sequence was
Guide-80'-80'-Open resulting in an energy resolution of $\Delta E\!  \sim\!0.1$
meV at the elastic line.  For the duration of the experiment, the sample was
mounted on a Cu mount and temperature was controlled in a $^3$He dilution
refrigerator. Measurements in an applied magnetic field were carried out on the
DCS time-of-flight spectrometer at NIST \cite{Copley:2003}.  All  measurements
on DCS were performed using a fixed incident neutron wavelength of
$\lambda_i=2.9$~\AA\@. The energy resolution on DCS was $\Delta
E\!\sim\!0.3$~meV at the elastic line. The sample was mounted in the (0,k,l)
scattering plane with (0,0,l) initially at 50$^{\circ}$ from the incident
neutron beam and then rotated through 120$^{\circ}$ in 0.5$^{\circ}$ steps
throughout the measurement. The sample was fixed on a Cu mount in a 11.5 T
vertical field cryomagnet with a dilution refrigerators insert. A magnetic
field between 4 and 11.5~T was applied along the a-axis and the sample was held
at T=100~mK for the duration of the measurements. Because of the very narrow
magnet aperture, measurements on DCS were confined to the $(0,k,l)$ scattering
plane with momentum transfers in the vertical direction limited to $h =
0\pm0.2$ r.l.u.\@.\

{\bf Author Contributions:} K.W.P. and Y.-J.K. conceived the experiments.
K.W.P., Y.Q., L.W.H. G.E.G, and A.I.K. performed the experiments and K.W.P.
analysed the data. C.R. provided additional data. K.H. and Y.B.K. developed the
theoretical model and performed calculations. G.J.S. and F.C.C. provided the
sample. K.W.P. and Y.-J.K. wrote the paper with contributions from all
co-authors.

%

\clearpage
\onecolumngrid
\appendix
\setcounter{figure}{0}
\setcounter{table}{0}
\renewcommand{\thefigure}{S\arabic{figure}}
\renewcommand{\thetable}{S\arabic{table}}
\begin{center}
{\large \bf Supplementay Information: Quasiparticle-continuum level repulsion 
    in a quantum magnet}
\end{center}
\section{Bond Operator Theory}
Considering the valence bond crystal order with valence bonds at the $J_4$
links in \BCPO\!\!, we have calculated triplon excitation spectrum within the
framework of bond operator theory as described below. To describe spin dynamics in \BCPO\!\!, we
employ a general spin Hamiltonian with Heisenberg $J_{ij}$,
Dzyaloshinskii-Moriya $\mathbf{D}_{ij}$, and anisotropic symmetric
$\Gamma_{ij}^{\mu\nu}$ exchange interactions
\begin{equation}
    \label{eq:Hamiltonian}
    \mathcal{H}\!=\!\sum_{i>j}\!\left(J_{ij}\mathbf{S}_i\!\cdot\!\mathbf{S}_j\!
            +\!\mathbf{D}_{ij}\!\cdot\!\mathbf{S}_i\!\times\!\mathbf{S}_j\!+\!\Gamma_{ij}^{\mu\nu}S^{\mu}_iS^{\nu}_j\right)-g\mu_B\mathbf{H}\cdot\!\sum_i\!\mathbf{S}_i,
\end{equation}
where summations over $\mu,\nu (=x,y,z)$ are assumed. The symmetric anisotropy
term is constrained by the relation
\begin{equation}
    \label{eq:Gamma}
    \Gamma_{ij}^{\mu\nu}\!=\!\frac{D_{ij}^\mu D_{ij}^\nu}{2J_{ij}} -
    \frac{\delta^{\mu\nu}\mathbf{D}_{ij}^2}{4J_{ij}}.
\end{equation}
Hence, for a given link, $J_{ij}$ and $\mathbf{D}_{ij}$ are regarded as free
parameters with $\Gamma_{ij}^{\mu\nu}$ determined by the former.
The coupling constants in the Hamiltonian are constrained with the space group
symmetry $Pnma$ into $J_n$, ${\bf D}_n$, and $\Gamma_n^{\mu\nu}$
$(n=1,\cdots,4)$ as introduced in main part of the paper. In our theory, the
anisotropic interactions ${\bf D}_3$, and $\Gamma_3^{\mu\nu}$ at $J_3$ links
are ignored since $J_3$ is already significantly small compared to the other
Heisenberg interactions.

Following standard procedures in bond operator theory
\cite{Sachdev:1990,Gopalan:1994,Matsumoto:2004}, we rewrite the Hamiltonian by
using the bond operator representations for the spins ${\bf S}_{L,R}$ in each
dimer at $J_4$ links
\begin{equation}
 S_{L,R}^{\alpha} = \pm \frac{1}{2} \left( s^{\dagger}t_{\alpha} + t_{\alpha}^{\dagger}s \right) - \frac{i}{2} \epsilon_{\alpha\beta\gamma}t_{\beta}^{\dagger}t_{\gamma},
\end{equation}
where $\alpha,\beta,\gamma \in \{x,y,z\}$ and $\epsilon_{\alpha\beta\gamma}$ is
the totally antisymmetric tensor. Here, the bond operators $s^{\dagger}$ and
$t_{\alpha}^{\dagger}$ create the spin-singlet and spin-triplet states at the
dimer respectively. We require the bosonic statistics for the bond operators.
The bond operator representation enlarges the Hilbert space so that we restrict
it to the physical Hilbert space by allowing only one bond particle at a dimer
via the Lagrange multiplier $\mu$: $-\mu (s^{\dagger}s +
t_{\alpha}^{\dagger}t_{\alpha}-1)$. On top of that, the $s$-bosons are
condensed to describe the valence bond crystal phase: $\langle s \rangle =
\langle s^{\dagger} \rangle = \bar{s}$.

After all above procedures, the bond operator Hamiltonian has the form $
\mathcal{H} = \mathcal{H}_{quad} + \mathcal{H}_{cubic} + \mathcal{H}_{quartic}
$, where $\mathcal{H}_{quad}$ consists of quadratic terms of $t$-boson
operators, and so forth. We consider the quadratic part $\mathcal{H}_{quad}$
for the description of single-triplon excitations in \BCPO\!\!. The quadratic
bond operator Hamiltonian is diagonalized through Bogoliubov transformation:
\begin{equation}
 \mathcal{H}_{quad}=E_{gr}+\sum_{\bf Q} \sum_{n=1}^6 \omega_n({\bf Q}) \gamma_{n}^{\dagger} ({\bf Q}) \gamma_{n} ({\bf Q}),
\end{equation}
where $\gamma_{n} ({\bf Q})$ is the Bogoliubov quasiparticle, here triplon,
operator with the excitation energy $\omega_n({\bf Q})$. Here, ${\bf Q}$ is
momentum and $n$ is band index of the triplon excitation spectrum.
The ground state is determined by the equations $ {\partial E_{gr}}/{\partial
    \bar{s}}={\partial E_{gr}}/{\partial \mu}=0 $.

As already discussed in main body of the paper, neutron scattering measurement
data is well reproduced in the bond operator theory with the following set of
coupling constants: $J_1\!=\!J_2\!=\!J_4\!=\!8$~meV, $J_3\!=\!1.6$~meV,
$D_1^a\!=\!0.6J_1$, $D_1^b\!=\!0.45J_1$, $\Gamma_1^{aa}\!=\!0.039J_1$,
$\Gamma_1^{bb}\!=\!-0.039J_1$, and
$\Gamma_1^{ab}\!=\!\Gamma_1^{ba}\!=\!0.135J_1$. With this set of coupling
constants, the ground state has $\bar{s}\!=\!0.869$, $\mu\!=\!-1.693J_1$.
Comparing with the case of an isolated dimer ($\bar{s}\!=\!1$,
$\mu\!=\!-0.75J$), we notice that the valence bonds in \BCPO are quite soft
with significant fluctuations, which is attributed to strong inter-dimer
interactions ($J_{1,2,3}$, $D_1$, $\Gamma_1$) comparable to intra-dimer
interaction ($J_4$).  An important comment is followed.  In the quadratic
Hamiltonian $\mathcal{H}_{quad}$, the DM terms with the coupling constant ${\bf
    D}_1$ cancel and do not explicitly enter the Hamiltonian, rather it is the
symmetric anisotropic interaction which is essential for describing magnetic
anisotropy in \BCPO\!\!.  Of course the existence of a large symmetric
anisotropy term necessarily implies a large antisymmetric DM term through
Eq.~\ref{eq:Gamma}.

The bond operator theory approach was able to successfully capture the magnetic
field dependence of triplon excitations for $H \parallel \mathbf{a}$ as
measured by neutron scattering. However, as an additional test of the theory we
may explore predictions for the Zeeman behaviour of the triplon bands with
magnetic fields applied along the other axes of the crystal. This can provide
information about the anisotropic critical fields which have been measured by
high field magnetization experiments~\cite{Kohama:12}.
\begin{figure}[htb]
    \centering
        \includegraphics{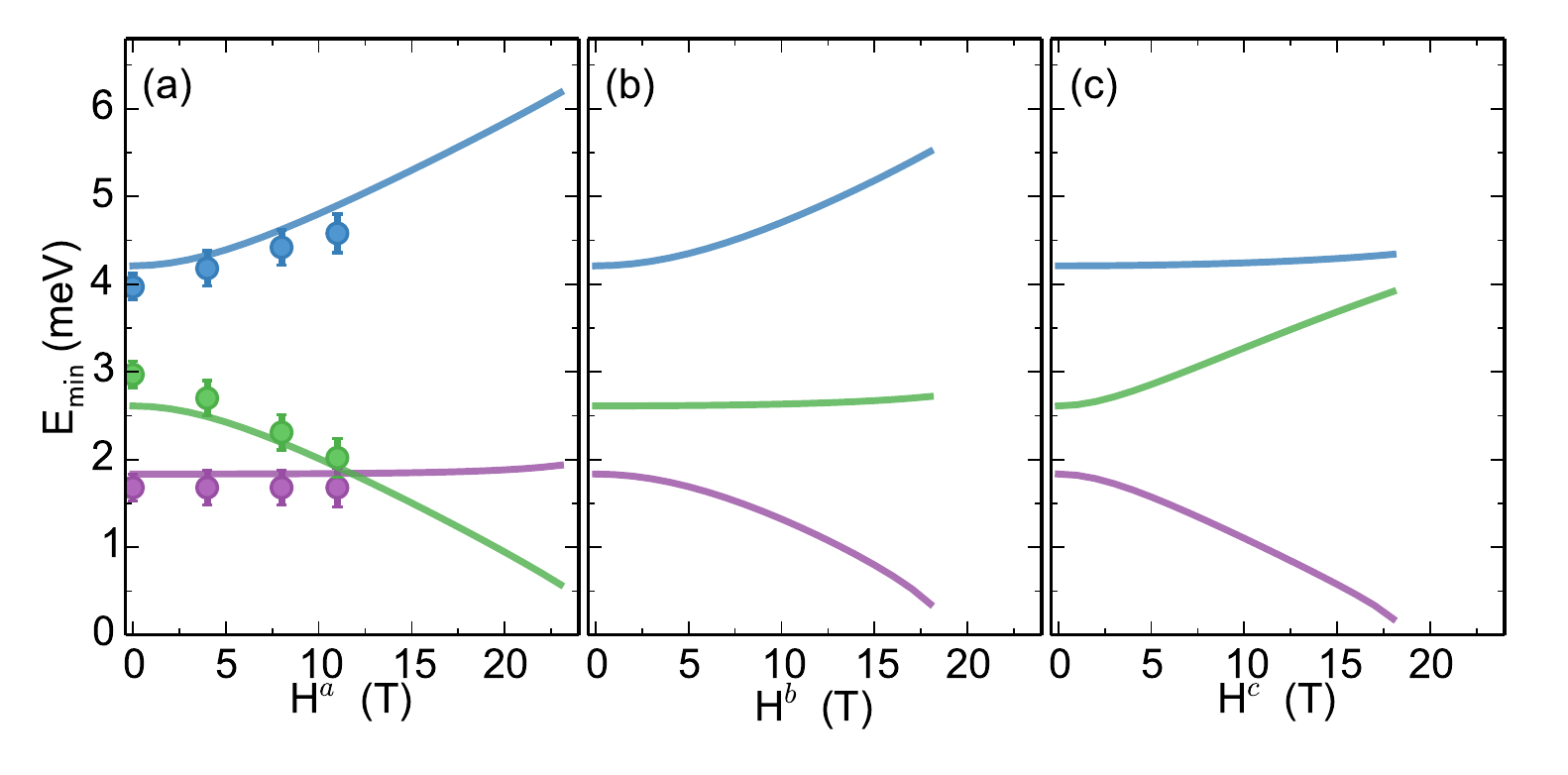}
        \caption{\label{fig:Zeeman} Calculated Zeeman energy of each mode at
            $\rm Q\!=\!(0,0.575,1)$ from the bond operator theory. In (a) the
            neutron scattering data for applied fields up to 11~T is included.}
\end{figure}
The Zeeman energy of the triplon bands at the dispersion minimum
$\mathbf{Q}\!=\!(0,0.575,1)$ calculated within the bond operator theory are
plotted in Fig.~\ref{fig:Zeeman}.  Each band has a characteristic field
direction, where the Zeeman dependence is flat, with negligible field
dependence.  The critical magnetic field $H_c$ at which the spin gap is closed
can be obtained by extrapolating the data points for each applied magnetic
field direction, giving $H_c^a\!\approx\!25$~T, $H_c^b\!\approx\!21$~T, and
$H_c^c\!\approx\!19$~T roughly in agreement with measured critical
fields~\cite{Kohama:12}. Besides the hierarchy in the critical fields, a rather
curious Zeeman behavior is revealed in Fig.~\ref{fig:Zeeman}
which shows the  magnetic field dependent energy of each
band is strongly coupled to the applied field direction with the modes having
the relative Zeeman character of $S_z =\{0,+1,-1\}$, $\{+1,0,-1\}$, and
$\{+1,-1,0\}$ for fields applied along the $\mathbf{a}$, $\mathbf{b}$, and
$\mathbf{c}$ axis respectively. Here, $S_z$ means the spin component along the
field direction.

\end{document}